\RequirePackage{fix-cm} 
\documentclass[a4paper, twoside, reqno, 12pt]{amsart}
\usepackage{fixltx2e}   

\usepackage{etex}

\usepackage[latin1]{inputenc}
\usepackage[T1]{fontenc}

\usepackage{esint}
\usepackage{dsfont}
\usepackage{xspace}
\usepackage{amsgen}
\usepackage{amsthm}
\usepackage{amssymb}
\usepackage{amsmath}
\usepackage{wasysym}
\usepackage{upgreek}
\usepackage{amsfonts}
\usepackage{stmaryrd}
\usepackage{mathtools}

\usepackage[mathcal, mathscr]{euscript}

\usepackage{mathrsfs}
\DeclareMathAlphabet{\mathscrbf}{OMS}{mdugm}{b}{n}

\usepackage{a4wide}

\usepackage[dvipsnames, table]{xcolor}
\definecolor{bckg}{RGB}{20.8, 20.8, 20.8}
\definecolor{oneblue}{rgb}{0.0, 0.0, 0.85}
\definecolor{Lightblue}{RGB}{214, 214, 214}
\definecolor{bluepigment}{rgb}{0.2, 0.2, 0.6}
\definecolor{charcoal}{rgb}{0.21, 0.27, 0.31}
\definecolor{denimblue}{rgb}{0.08, 0.38, 0.74}
\definecolor{Lightgray}{rgb}{0.89, 0.89, 0.89}
\definecolor{darkgrey}{rgb}{0.273, 0.281, 0.30}
\definecolor{darkelectricblue}{rgb}{0.33, 0.41, 0.47}

\usepackage[sort&compress, comma, square, numbers]{natbib}

\usepackage{booktabs}

\usepackage{psfrag}
\usepackage{graphicx}
\usepackage{subfigure}
\usepackage{morefloats}
\usepackage{indentfirst}

\usepackage{acronym}
\usepackage{microtype}

\usepackage[perpage, symbol]{footmisc}

\usepackage[usenames, dvipsnames, pdf]{pstricks}
\usepackage{epsfig}
\usepackage{pst-grad} 
\usepackage{pst-plot} 

\usepackage[colorlinks,
           urlcolor=oneblue,
           linkcolor=denimblue,
           citecolor=NavyBlue,
           bookmarksopen=false,
           pdfpagemode=UseNone,
           pagebackref]{hyperref}

\usepackage[labelsep=period,%
            labelfont={bf,sf,color=bluepigment},%
            justification=raggedright]{caption}

\usepackage[explicit]{titlesec}

\titleformat{\section}[block]
  {\color{NavyBlue}\Large\sffamily\bfseries}
  {}
  {0.0em}
  {\colorbox{bckg!5}{\strut\parbox{\dimexpr\linewidth-2\fboxsep\relax}{\thesection. #1}}}
  [\vspace*{0.33em}]

\titleformat{name=\section,numberless}[block]
  {\color{NavyBlue}\Large\sffamily\bfseries}
  {}
  {0.0em}
  {\colorbox{bckg!5}{\strut\parbox{\dimexpr\linewidth-2\fboxsep\relax}{#1}}}
  [\vspace*{0.33em}]

\titleformat{\subsection}
  {\color{NavyBlue}\large\sffamily\bfseries}
  {}
  {0.0em}
  {\colorbox{bckg!5}{\parbox{\dimexpr\linewidth-2\fboxsep\relax}{\thesubsection. #1}}}
  [\vspace*{0.33em}]

\titleformat{name=\subsection,numberless}
  {\color{NavyBlue}\Large\sffamily\bfseries}
  {}
  {0em}
  {\colorbox{bckg!5}{\parbox{\dimexpr\linewidth-2\fboxsep\relax}{#1}}}
  [\vspace*{0.33em}]

\titleformat{\subsubsection}
  {\color{bluepigment}\sffamily\normalsize\bfseries}
  {\thesubsubsection}
  {0.5em}
  {#1}
  [\vspace*{0.33em}]

\titleformat{\paragraph}[runin]
  {\color{bluepigment}\sffamily\small\bfseries}
  {}
  {0em}
  {#1}

\titlespacing{\section}{0.0em}{1.5em plus 2pt minus 2pt}%
{1.0em plus 2pt minus 2pt}[0em]
\titlespacing{\subsection}{0.5em}{1.5em plus 2pt minus 2pt}%
{1.0em}[0em]
\titlespacing{\subsubsection}{0.5em}{1.5em plus 2pt minus 2pt}%
{1.0em plus 2pt minus 2pt}[0em]

\usepackage{titletoc}

\contentsmargin{0.5em}
\newlength{\tocsep} 
\titlecontents{section}[\tocsep]
  {\addvspace{10pt}\bfseries\sffamily}
  {\contentslabel[\thecontentslabel]{\tocsep}}
  {}
  {\ \titlerule*[0.75pc]{.}\ \thecontentspage}
  []
\titlecontents{subsection}[\tocsep]
  {\addvspace{8pt}\sffamily}
  {\contentslabel[\thecontentslabel]{\tocsep}}
  {}
  {\ \titlerule*[0.5pc]{.}\ \thecontentspage}
  []
\titlecontents*{subsubsection}[\tocsep]
  {\addvspace{2pt}\footnotesize\sffamily}
  {}
  {}
  {\ \titlerule*[0.35pc]{.}\ \thecontentspage}
  [\\*]

\makeatletter
\def\@setauthors{%
  \begingroup
  \def\thanks{\protect\thanks@warning}%
  \trivlist
  \centering\footnotesize \@topsep30\p@\relax
  \advance\@topsep by -\baselineskip
  \item\relax
  \author@andify\authors
  \def\\{\protect\linebreak}%
  \textsc{\normalsize\textcolor{darkelectricblue}{\authors}}%
  \ifx\@empty\contribs
  \else
    ,\penalty-3 \space \@setcontribs
    \@closetoccontribs
  \fi
  \endtrivlist
  \endgroup
}
\def\@settitle{\begin{center}%
  \baselineskip14\p@\relax
    \bfseries
    \textsc{\Large\textcolor{charcoal}{\@title}}
  \end{center}%
}
\makeatother

\usepackage{enumitem}
\setlist[description]{%
  topsep=30pt,               
  itemsep=5pt,               
  font={\bfseries\sffamily\color{NavyBlue}}, 
}

\usepackage{fancyhdr}
\usepackage{lastpage}

\newcommand*\Title{\textcolor{bluepigment}{Coupling conditions for shallow water equations}}
\newcommand*\Authors{\textcolor{bluepigment}{J.-G.~Caputo, D.~Dutykh \& B.~Gleyse}}
\newcommand*{\plogo}{\textcolor{gray}{{\texttt{arXiv.org} / \textsc{hal}}}} 

\numberwithin{equation}{section}

\newcommand{\be}{\begin{equation}}
\newcommand{\ee}{\end{equation}}

\usepackage{environ}
\NewEnviron{myequation}{%
\begin{equation}
\scalebox{0.88}{$\BODY$}
\end{equation}}

\renewcommand{\alpha}{\upalpha}

\renewcommand{\gamma}{\boldsymbol{\upgamma}}

\usepackage{acronym}

\begin{document}

\title[\Title]{Coupling Conditions for Water Waves at Forks}

\author[J.-G.~Caputo]{Jean-Guy Caputo}
\address{\textbf{J.-G.~Caputo:} Laboratoire de Math\'ematiques, INSA de Rouen, BP 8, Avenue de l'Universit\'e, Saint-Etienne du Rouvray, 76801 France}
\email{caputo@insa-rouen.fr}
\urladdr{https://sites.google.com/site/jeanguycaputo/}

\author[D.~Dutykh]{Denys Dutykh$^*$}
\address{\textbf{D.~Dutykh:} Univ. Grenoble Alpes, Univ. Savoie Mont Blanc, CNRS, LAMA, 73000 Chamb\'ery, France}
\email{Denys.Dutykh@univ-savoie.fr}
\urladdr{http://www.lama.univ-savoie.fr/~dutykh/}
\thanks{$^*$ Corresponding author}

\author[B.~Gleyse]{Bernard Gleyse}
\address{\textbf{B.~Gleyse:} Laboratoire de Math\'ematiques, INSA de Rouen, BP 8, Avenue de l'Universit\'e, Saint-Etienne du Rouvray, 76801 France}
\email{gleyse@insa-rouen.fr}

\keywords{Networks; nonlinear shallow water equations; nonlinear wave equations}


\begin{titlepage}
\setcounter{page}{1}
\thispagestyle{empty} 
\noindent
{\Large Jean-Guy \textsc{Caputo}}\\
{\it\textcolor{gray}{INSA de Rouen, France}}\\[0.02\textheight]
{\Large Denys \textsc{Dutykh}}\\
{\it\textcolor{gray}{CNRS, Universit\'e Savoie Mont Blanc, France}}\\[0.02\textheight]
{\Large Bernard \textsc{Gleyse}}\\
{\it\textcolor{gray}{INSA de Rouen, France}}\\[0.16\textheight]

\colorbox{Lightblue}{
  \parbox[t]{1.0\textwidth}{
    \centering\huge\sc
    \vspace*{0.7cm}
    
    \textcolor{bluepigment}{Coupling Conditions for Water Waves at Forks}

    \vspace*{0.7cm}
  }
}

\vfill 

\raggedleft     
{\large \plogo} 
\end{titlepage}


\newpage
\thispagestyle{empty} 
\par\vspace*{\fill}   
\begin{flushright} 
{\textcolor{denimblue}{\textsc{Last modified:}} \today}
\end{flushright}


\newpage
\maketitle
\thispagestyle{empty}


\begin{abstract}

We considered the propagation of nonlinear shallow water waves in a narrow channel presenting a fork. We~aimed at computing the coupling conditions for a 1D effective model, using~2D simulations and an analysis based on the conservation laws. For small amplitudes, this analysis justifies the well-known Stoker interface conditions, so~that the coupling does not depend on the angle of the fork. We~also find this in the numerical solution. Large amplitude solutions in a symmetric fork also tend to follow Stoker's relations, due to the symmetry constraint. For~non symmetric forks, 2D~effects dominate so that it is necessary to understand the flow inside the fork. However, even~then, conservation laws give some insight in the dynamics.


\bigskip
\noindent \textbf{\keywordsname:} Networks; nonlinear shallow water equations; nonlinear wave equations \\

\smallskip
\noindent \textbf{MSC:} \subjclass[2010]{ 35Qxx (primary), 35R02 (secondary)}
\smallskip \\
\noindent \textbf{PACS:} \subjclass[2010]{ 47.35.Bb (primary), 47.15.gm (secondary)}

\end{abstract}


\newpage
\tableofcontents
\thispagestyle{empty}


\newpage
\section{Introduction}

The propagation of nonlinear waves in a network is an important topic. As an example, consider a hydrological network which is prone to floods. Understanding the global dynamics of the network can help identify its most vulnerable sections and take the appropriate measures. Real networks are formed by long 2D or 3D channels of a small cross-section. To study the propagation of waves in such systems, a first step is to consider a simple fork as a model of elementary junctions. The~final goal is to reduce the model to 1D channels connected by appropriate interface conditions. The~study of such 1D systems is now well advanced, in particular for systems of conservation laws, see the review~\cite{b14}.

The type of PDE model describing the quantity propagating on the network is very important to drive the coupling conditions. Recently for the sine-Gordon nonlinear wave equation, we~\cite{cd14} introduced a homothetic reduction~\cite{homothetie} where we averaged the operator over the fork region and consistently took the limit when the width tended to zero. Assuming continuity of the field, we obtained Kirchhoff's law for the gradients. Comparing the 2D solution with the one for the reduced 1D equations gives excellent agreement. In this situation, the angle of the fork does not play a role. When considering networks of rivers, many authors, for example Stoker~\cite{Stoker1958b} and Jacovkis~\cite{jacovkis} assumed continuity of the water height and continuity of the flux so that the angle of the fork did not come in. In the close context of gas dynamics, Holden and Risebro~\cite{holden} studied shocks in a pipe with an elbow. They showed that the Riemann problem had a unique solution when the angle was smaller than $\pi$. For classical hydrodynamics, the angle is important, in a fork, it sets the forces experienced by the pipes~\cite{Landau}. In fact, for large amplitude shallow water waves our numerical calculations show that the energy entering a branch can vary from 20\% to 50\% depending on the symmetry of the fork. These~studies point out the importance of the angle.

A few authors addressed the problem of the angle of a junction. Schmidt~\cite{Schmidt} studied the 2D connection between 1D channels; he made no assumption on the size of the connecting domain. The~flow in the junction was assumed linear so that the author used a variational method that gave the solution as a superposition of fields. The~final result was a system of ordinary differential equations for the values at the ends of the branches coupled to the shallow water PDEs. Despite its formal beauty, it~remains difficult to handle and does not give a simple picture. Shi et al.~\cite{shi05} studied experimentally and numerically the propagation of long waves in wide and narrow channels. They~used the Boussinesq dispersive shallow water equations for narrow channels. They observed no angle dependence and a strong transmission. For the same equations, Nachbin and Simoes~\cite{ns15} obtained interface conditions containing implicitly the angles of the fork. These gave an excellent matching between the average of the 2D solution and the solution of the 1D effective model for angles smaller than $\pi/3$.

In this article, we consider the nonlinear shallow water equations. The~system is very general because it only involves conservation laws. Also it is simple enough. We revisit the problem of shallow water propagation in 2D forks using our homothetic reduction procedure to obtain approximate conservation laws and compare them with the numerical solutions. We compute approximate conservation for the mass, momenta and energy laws for a general fork geometry. In the small amplitude limit we recover Stoker's conditions, i.e., continuity of surface elevation and mass conservation (Kirchoff law). To our knowledge, this is a first formal justification of Stoker's interface conditions. This angle independent reduction holds also for a general class of scalar nonlinear wave equations, for example the 2D sine-Gordon equation or the 2D reaction-diffusion equation; it~confirms the results of~\cite{cd14}. We computed the 2D numerical solution for a simple T-fork geometry for small and large amplitudes. The~wave was also launched in two different branches to see the effect of symmetry. We show that Stoker's conditions hold for the symmetric case for small and large amplitudes. For the non-symmetric case, they hold for small amplitudes. When the amplitude is large, 2D effects dominate the fork region. Nevertheless the approximate conservation laws give an insight into the flow.

The article is organized as follows. Section~\ref{sec2} presents the fork geometry and shows the straightforward reduction for a general class of nonlinear wave equations. In Section~\ref{sec3} we recall the shallow water equations and their conserved quantities. Section~\ref{sec4} gives the integrals of these equations on the fork showing that the mass and energy laws do not involve the angles while the momenta laws do. Section~\ref{sec5} shows the 2D numerical solutions for symmetric and non symmetric configurations for small and large waves. There, we compare the numerical results with the conservation laws established in Section~\ref{sec4}. We discuss these results and conclude in Section~\ref{sec6}.


\section{General Scalar Nonlinear Wave Equations} \label{sec2}

Before considering the nonlinear shallow water equations, we analyze the simpler case of a class of scalar 2D nonlinear wave equations. This large class includes hyperbolic wave equations like the sine-Gordon equation as well as reaction diffusion equations like the Fisher equation, to name a few. We consider equations of the form
\be \label{nlwave} \alpha u_{tt} + \beta u_{t} - \Delta u = N(u), \ee
where $u(x,y,t)$ is a scalar, $\Delta$ is the usual 2D Laplacian and where $N(u)$ is a nonlinearity not containing derivatives. The boundary condition on the lateral domain is of Neumann type 
\be \label{bc_nlwave} \partial_n u= \nabla u \cdot \mathbf{n}=0.\ee

\begin{figure}
  \centering
  \includegraphics[scale=0.48]{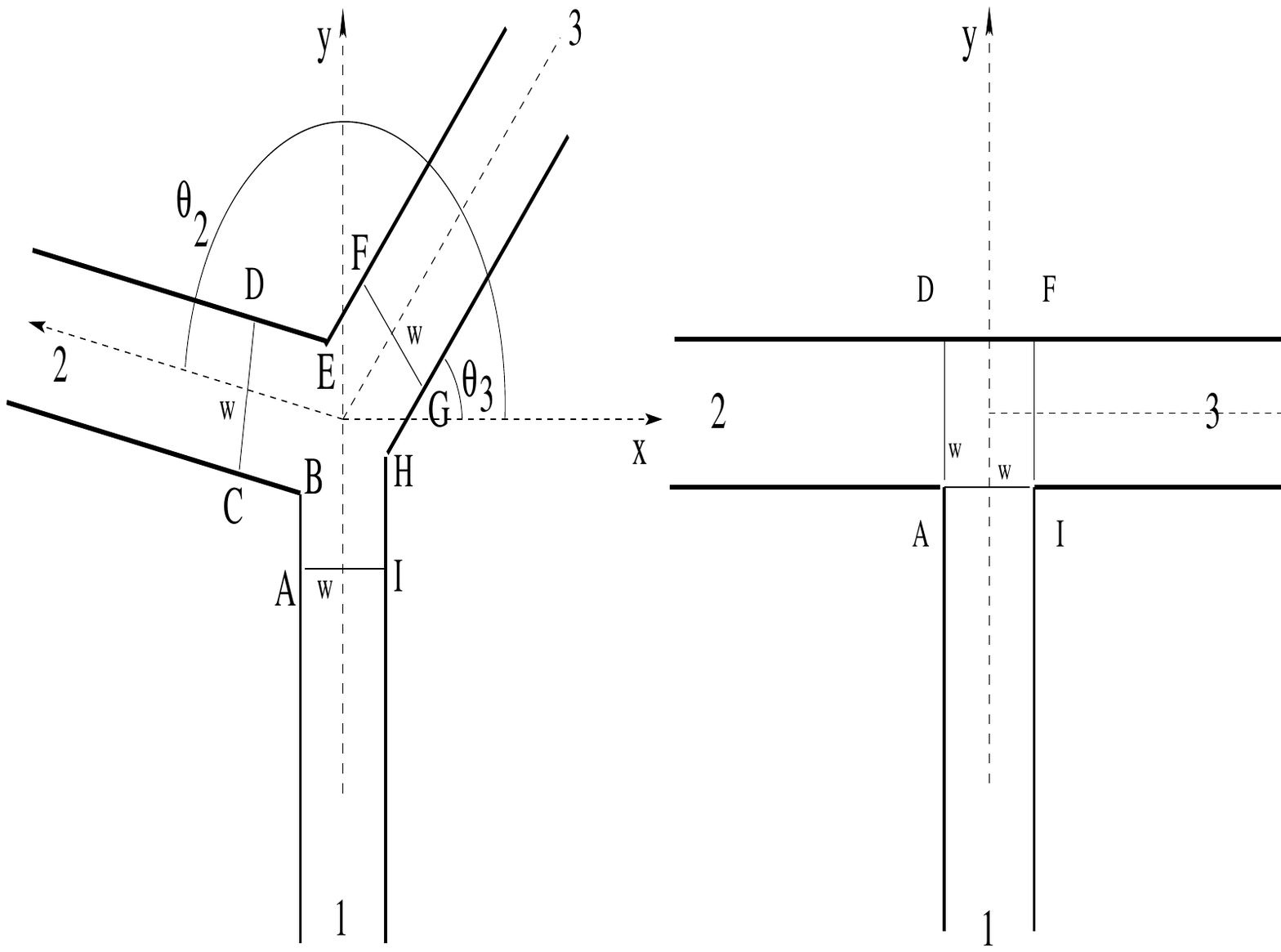}
  \caption{A fork geometry with arbitrary angles (\textbf{left}) and with right angles (\textbf{right}).}
  \label{elbow2}
\end{figure}

Consider the fork domain shown in Figure~\ref{elbow2}. Far from the fork region, the solution can be assumed to be 1D so that we do not loose much information by approximating the 2D dynamics with a 1D equation. Inside the fork domain, a strong coupling occurs between the branches. To see this, we proceed as in~\cite{cd14} and integrate the operators on the fork region. Then we examine the behavior of the different terms as $w$, the width of the branches, goes to zero. We assume that domains that we consider behave in a regular way as we shrink $w$ homothetically to zero,~\cite{homothetie}.

Consider the asymmetric Y-branch shown in the left panel of Figure~\ref{elbow2}. A first assumption is the continuity of $u$ which is obvious for the 2D operator. The other condition comes from the integration of the operator (\ref{nlwave}) on the fork domain $\mathcal{F}=IABCDEFGHI$. We get
\be \label{flux} \int [  \alpha u_{tt} + \beta u_{t} -N(u) ] ~  dx dy
- \int_{\partial \mathcal{F}}  (\nabla u) \cdot \mathbf{n}~ ds = 0 .\ee

The first integral is of order $O(w^2)$. On the exterior boundaries, $(\nabla u) \cdot \mathbf{n} = 0$ so the line integral reduces to 
$$ \int_{IA} \dots + \int_{CD} \dots +\int_{FG} \dots\, ,$$ 
which are $O(w)$. We then obtain for $w\to 0$
\be\label{jump_nlwave} 
- \partial_s u_1 + \partial_s u_2 + \partial_s u_3 =0 , \ee
where $u_i,~i=1,2,3$ are respectively the values of the field at the end of branch 1 (IA) and at the beginning of branches 2 (FG) and 3 (CD). Relation (\ref{jump_nlwave}) is Kirchhoff's law~\cite{cd14}. When the widths of the branches are not equal, this Kirchoff relation becomes 
\be\label{jump_nlwave2}
- w_1 \partial_s u_1 + w_2 \partial_s u_2 + w_3 \partial_s u_3 =0 .\ee

Remark that in the result \eqref{jump_nlwave} the angle of the fork
plays no role. The reduction leading from the flux equation to (\ref{jump_nlwave2}) is an asymptotic result that holds for $w \to 0$. It is then natural to approximate the 2D equation \eqref{nlwave} by a 1D equation in each branch together with the conditions of continuity and Kirchoff \eqref{jump_nlwave} at the junctions.

The result we obtain can be connected to a property of the Laplace operator with Neumann boundary conditions on a so-called ``fat'' graph~\cite{exner08}. Consider a graph where each edge has a transverse size $w$, assume Neumann boundary conditions on the transverse edge. Then the spectrum of the Laplacian converges to the one of the 1D Laplacian as $w \to 0$. This is true for compact and non compact graphs. See the article by Exner and Post~\cite{exner08} and the book by Post~\cite{post12} for the details of the proof.

\begin{figure}
\centerline{\resizebox{9 cm}{6 cm}{\includegraphics{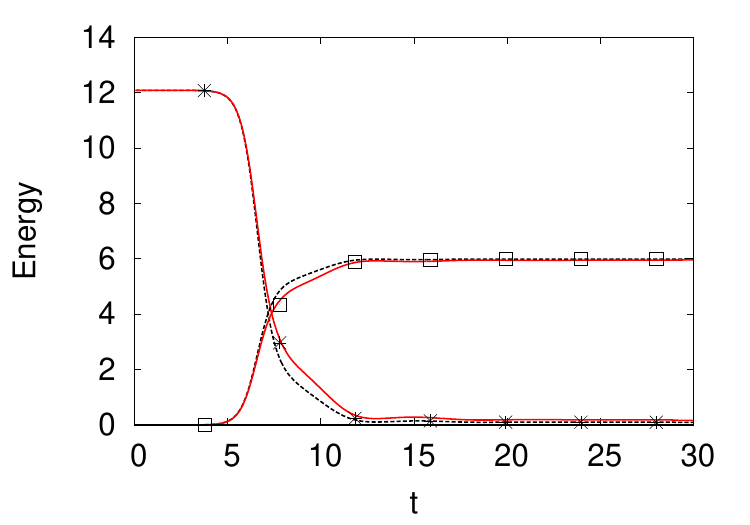}}}
\caption{Time evolution of the energies $E_2^i$ for the kink motion in branches $i=1,2$ for the T-junction (90 degrees) in full line (red online), for the Y-junction (45 degrees) in dashed line. The energy $E_1^i$ for the 1D effective model is plotted with points.}
\label{ekink90_45}
\end{figure}

The validity of the reduction was confirmed numerically for the 2D sine-Gordon equation, \eqref{nlwave} with $\alpha=1, ~\beta=0$ and $N(u)=-\sin(u)$ in~\cite{cd14}. There we compared the 2D solutions to the ones of the 1D sine-Gordon equation in each branch, coupled by the interface conditions. For completeness, we recall the case of a sine-Gordon kink propagating in forks with angles $45$ and $90$ degrees. The kink is an exact solution in 1D, it is
\be\label{kink}
  u(x,t) = 4\arctan\left[\exp({ x - vt \over \sqrt{1 - v^2}})\right],
\ee
where the velocity $0 \leq v < 1$ is a free parameter. To compare the 2D and 1D solutions, we plot the energies in each branch
\be\label{enk2}
E_2^i = \int_{\Omega_i} \left[{1\over 2}u_t^2 + {1\over 2}|\nabla u|^2
+ (1 - \cos u)\right]\, d x d y , \ee
and 
\be\label{enk1}
E_1^i = \sum_{i=1,2,3} \int_{\Omega_i}
\left[{1\over 2}u_t^2 + {1\over 2}|u_x|^2
+ (1 - \cos u)\right]\, d x , \ee
where $\Omega_i$ is branch $i$, abusively named the same in 1D and 2D. The kink is started in branch 1 with an initial velocity $v=0.75$, this gives a typical wavelength $\lambda \approx 4/\sqrt{1-v^2}= 2.7$. The width of the branches is $w=0.7 \ll \lambda$. Figure~\ref{ekink90_45} shows the time evolution of the energies $E_2^i$ for forks with angles 45 and 90 degrees and $E_1^i$, where $i=1,2$ corresponds to the branches. Initially the kink is in branch 1 so that $E_2^2=E_2^3 =0$. As the kink crosses into branches 2 and 3, $E_2^1$ becomes very small. Note the excellent agreement between the two expressions $E_2^i$ and the expression $E_1^i$. This confirms that the angle of the fork plays no role for such a system.

The dynamics of kinks for the sine-Gordon equation is controlled by the energy: if the initial energy is enough, a kink in branch 1 gives rise to two kinks in branches 2 and 3. This gives a very simple picture. Other solutions like the breather have much more complicated dynamics, we refer the reader to~\cite{cd14} for more details.


\section{The Nonlinear Shallow Water Equations} \label{sec3}

The shallow water equations in a 2D domain written in terms of the fluid velocity $\mathbf{u}(x,t)$
$$\mathbf{u} = (u , v)^T$$
and the water height $h(x,t)$ read~\cite{Stoker1958b}
\begin{eqnarray} 
  h_t + \nabla \cdot (h \mathbf{u}) =0,  \label{sha1}\\
  (h u)_t  + \nabla  \cdot \left ( \begin{matrix} h u^2 + {g h^2 \over 2} \cr h u v \end{matrix} \right ) =0, \label{sha2a} \\
  (h v)_t  + \nabla  \cdot \left ( \begin{matrix} h u v \cr h v^2 + {g h^2 \over 2} \end{matrix} \right ) =0 , \label{sha2b} 
\end{eqnarray}
where $g$ is the gravitational acceleration. The wall boundary condition is 
\be \label{bc_sha} \mathbf{u} \cdot \mathbf{n} = 0 .\ee

We assume an even bottom of the channels $h=h_0$.


\subsection{Conserved Quantities}

We first recall the conserved quantities. Integrating Equations (\ref{sha1})--(\ref{sha2b}) over a 2D closed domain $\Omega$ and using the boundary condition (\ref{bc_sha}) we get
\begin{equation}
  \partial_t \int_\Omega h~ dx dy=0, \label{isha1}
\end{equation}
\begin{equation}
  \partial_t \int_\Omega h u ~  dx dy + \oint_{\partial \Omega} {g h^2 \over 2} {n}_x~ ds =0, \label{isha2a} 
\end{equation}
\begin{equation}
  \partial_t \int_\Omega h v ~  dx dy + \oint_{\partial \Omega} {g h^2 \over 2} {n}_y~ ds =0.   \label{isha2b}
\end{equation}

A localized wave will have as first conserved quantity the integral of the water elevation
$$ M = \int_\Omega h~ dx dy  . $$

The total x and y momenta 
$$P_x = \int_\Omega h u ~ dx dy , ~~ ~~ P_y = \int_\Omega h v~ dx dy $$
will not be conserved in the fork geometries.

A flux relation that can be deduced from the conservation laws 
(\ref{sha1})--(\ref{sha2b}) is the total energy flux 
\be\label{en_flux}
  e_t + \nabla \cdot \left [ \mathbf{u}(e + {gh^2 \over 2})\right ] =0. \ee
where the total energy density~is
\be\label{en_sh} e = {1 \over 2} \left [  g h^2 + ( u^2  + v^2 ) h \right ] . \ee

Integrating the energy flux  relation over a volume $\Omega$ we obtain that a localized wave in $\Omega$ will have constant energy
$$ {d E \over dt} = {d \over dt} \int_\Omega e ~ dx dy =0.  $$


\subsection{Small Amplitude Limit}

It is well known that in the linear limit, Equations (\ref{sha1})--(\ref{sha2b}) reduce to the linear wave equation for the water height $h$. To see this, consider the steady state $h=h_0, ~u=v=0$, then the linearized system~is
\begin{eqnarray} 
  h_t + h_0 \nabla \mathbf{u}=0,  \label{lsha1}\\
  h_0 \mathbf{u}_t + g \nabla h =0 , \label{lsha2a} 
\end{eqnarray}
The boundary conditions reduce to $\nabla h \cdot \mathbf{n}=0$ as can be seen
by projecting (\ref{lsha2a}) on $\mathbf{n}$. This equation is in the class
(\ref{nlwave}).


\section{Reduction of the Shallow Water Equations}\label{sec4}

The shallow water equations cannot be reduced so simply as the nonlinear scalar wave equation. In fact, it is not clear what are the right interface conditions that should be implemented for a 1D effective model. Stoker, in his well-known book~\cite{Stoker1958b} introduces the following interface conditions for the water elevations $h_1,h_2,h_3$ and branch-oriented velocities $u_1^{\parallel}, u_2^{\parallel}, u_3^{\parallel}$
\begin{eqnarray} \label{stoker2}
  h_1 = h_2=h_3, \\
  - h_1 u_1^{\parallel}  +  h_2 u_2^{\parallel} + h_3 u_3^{\parallel} =0,
\end{eqnarray}
and uses them to analyze the junction of the Mississippi and the Missouri rivers. These conditions were not justified by a formal argument. Note also that they do not depend on the angle of the junction. 

Below, we will see that these conditions arise naturally in the limit of small amplitude for the shallow water equations. For general amplitudes, it is not clear that these apply. To analyze the problem, we proceed as in~\cite{cd14}, integrate the governing equations on the bifurcation region and consider the limit of vanishing transverse width $w$.


\subsection{Mass Flux}

Integrating the Equation~(\ref{sha1}) over the closed region $\mathcal{F}\equiv ABCDEFGHIA$ yields
$$\int_{\mathcal{F}} h_t ~dx dy + \oint_{\partial \mathcal{F}} h~ \mathbf{u} \cdot \mathbf{n}~ ds =0.$$

Because of the boundary condition $\mathbf{u} \cdot \mathbf{n} =0$ on ABC, DEF and GHI the expression above reduces to
$$\int_{\mathcal{F}}   h_t~dx dy + \int_{AI}  h~ \mathbf{u} \cdot \mathbf{n} ~ds
+ \int_{CD}  h~ \mathbf{u} \cdot \mathbf{n} ~ds 
+ \int_{FG}  h~ \mathbf{u} \cdot \mathbf{n} ~ds = 0. $$

The first integral is $O(w^2)$ while the three other integrals are $O(w)$. Dividing the equation by $w$ and taking the limit $w\to 0$ we get from these three terms
\be\label{red_sha1}
- h_1 u_1^{\parallel} + h_2 u_2^{\parallel} + h_3 u_3^{\parallel} = 0,
\ee
where we have introduced the local branch-oriented velocities $u^{\parallel}, u^{\perp}$ such that
\be\label{ubranch}\begin{pmatrix} u^{\parallel} \\ u^{\bot} \end{pmatrix} = 
\begin{pmatrix} \cos \theta  & \sin \theta \\ -\sin \theta & \cos \theta \end{pmatrix} \begin{pmatrix} u \\ v \end{pmatrix} \ee
and where the indices 1,2 and 3 refer to the branches. Of course, when the transverse widths $w_1, w_2, w_3$ are different, with the condition that the ratios $w_2/w_1, w_3/w_1$ remain finite, the relation (\ref{red_sha1}) becomes
$$- w_1 h_1 u_1^{\parallel} +  w_2 h_2 u_2^{\parallel} +  w_3 h_3 u_3^{\parallel}=0.$$


\subsection{Energy Flux}

The energy flux (\ref{en_flux}) can be consistently reduced to a 1D relation. As for the mass relation, we integrate Equation~(\ref{en_sh}) over the domain $\mathcal{F}=$ ABCDEFGHIA to obtain
$$\int_{\mathcal{F}} e_t ~dx dy + 
\oint_{\partial \mathcal{F}} (e + {gh^2 \over 2})~ \mathbf{u} \cdot \mathbf{n} ~ds =0.$$

Because of the boundary condition $\mathbf{u} \cdot \mathbf{n} =0$
on ABE, the expression above reduces to
$$\int_{\mathcal{F}} e_t ~dx dy
+ \int_{AI}  (e + {gh^2 \over 2})~ \mathbf{u} \cdot \mathbf{n} ~ds
+ \int_{CD} (e + {gh^2 \over 2})~ \mathbf{u} \cdot \mathbf{n} ~ds 
+ \int_{FG} (e + {gh^2 \over 2})~ \mathbf{u} \cdot \mathbf{n} ~ds 
= 0 . $$

The first integral is $O(w^2)$ while the three other integrals are $O(w)$. Dividing the equation by $w$ and taking the limit $w\to 0$ we get from these three terms
\be\label{red_en_fork}
- (e_1 + {g h_1^2 \over 2})   u_1^{\parallel} 
+  (e_2 + {g h_2^2 \over 2}) u_2^{\parallel} 
+  (e_3 + {g h_3^2 \over 2}) u_3^{\parallel} 
=0 .\ee

To conclude, Equation~(\ref{sha1}) gives in the 1D limit, the balance of mass (\ref{red_sha1}). The~same happens for the energy flux (\ref{en_flux}) which yields (\ref{red_en_fork}). The natural matching conditions for 1D shallow water equations on a network are then
\be \label{mass}
- h_1 u_1^{\parallel}  +  h_2 u_2^{\parallel} + h_3 u_3^{\parallel} =0,\ee
\be\label{energy}
- u_1^{\parallel} ({g h_1^2} + h_1{{u_1^{\parallel}}^2 \over 2}) 
+ u_2^{\parallel} ({g h_2^2} + h_2 {{u_2^{\parallel}}^2 \over 2}) 
+ u_3^{\parallel} ({g h_3^2} + h_3 {{u_3^{\parallel}}^2 \over 2}) =0 .
\ee

For the mass and the energy balance laws, we have a similar situation to the one of the nonlinear scalar wave equation, the angles of the fork do not play any role. In the small amplitude limit, the speeds $u_1,u_2,u_3$ are small and the squares can be neglected in the energy relation. Then, we recover the Stoker interface conditions (\ref{stoker2}).


\subsection{Momentum Flux for a General Fork}

Contrary to the mass and the energy, the momentum Equations (\ref{sha2a})--(\ref{sha2b}) cannot be consistently reduced to a 1D condition involving $h,u^{\parallel}$ at each end of $\mathcal{F}$. 

To see this, integrate the horizontal momentum Equation~(\ref{sha2a}) over the 
domain $\mathcal{F}$ and get
$$ \int_{\mathcal{F}}  (h u)_t ~dx dy + \oint_{\partial\mathcal{F}} 
\left ( \begin{matrix} h u^2 + {g h^2 \over 2} \cr h u v \end{matrix} \right ) 
\cdot \mathbf{n} ~ds  =0 , $$
where the first integral is a surface integral and the second one a line integral. In the integrand of the latter, we have 
$$ \left (\begin{matrix} h u^2  \cr h u v \end{matrix} \right ) 
\cdot \mathbf{n}=
h u \left (\begin{matrix}  u  \cr v \end{matrix} \right ) \cdot \mathbf{n}= 0$$
on the exterior boundaries of ${\partial\mathcal{F}} $ because of the boundary condition (\ref{bc_nlwave}). Then, only the potential term ${g h^2 \over 2}$ will contribute to these terms.

The $O(w)$ terms (line integrals) reduce to
\begin{myequation}
\begin{array}{c}
-{g \over 2}(|AB| h_{AB}^2  - |HI| h_{HI}^2 )
-\sin \theta_2 {g \over 2} (|BC| h_{BC}^2 -|DE|  h_{DE}^2 )
- \sin \theta_3 {g \over 2}(|EF| h_{EF}^2 -|HG| h_{HG}^2)\\
-w h_1 u_1 v_1 
+ w \left [ (h_2 u_2^2 + g {h_2^2 \over 2}) \cos \theta_2 + h_2 u_2 v_2 \sin \theta_2 \right ]  
+ w \left [ (h_3 u_3^2 + g {h_3^2 \over 2}) \cos \theta_3 + h_3 u_3 v_3 \sin \theta_3 \right ]  =0.
\end{array}
\label{cpx}
\end{myequation}

Using the branch oriented velocities (\ref{ubranch})
we get the approximate law 
\begin{equation}
\begin{array}{c}
-{g \over 2}(|AB| h_{AB}^2  - |HI| h_{HI}^2 )
-\sin \theta_2 {g \over 2} (|BC| h_{BC}^2 -|DE|  h_{DE}^2 )
- \sin \theta_3 {g \over 2}(|EF| h_{EF}^2 -|GH| h_{GH}^2)\\
-w h_1 u_1 v_1 
+ w~ \cos \theta_2 \left [h_2 {u_2^{\parallel}}^2+ g {{h_2}^2 \over 2} \right ]
+ w~ \cos \theta_3 \left [h_3 {u_3^{\parallel}}^2+ g {{h_3}^2 \over 2} \right ]
=0 ,
\end{array}
\label{cpx2}
\end{equation}
where we neglected the velocity components $u^{\perp}$.

Similarly for the vertical momentum equation we obtain
$$ {g \over 2} \cos \theta_2 (|BC|~h_{BC}^2 -|DE| ~h_{DE}^2 )
+  {g \over 2} \cos \theta_3 (|EF|~h_{EF}^2 -|GH| ~h_{GH}^2 )
-w ~ \left [h_1 v_1^2 + g {{h_1}^2 \over 2} \right ]$$
\be  \label{cpy}
+ w \left [ (h_2 u_2^2 + g {h_2^2 \over 2}) \sin \theta_2 + h_2 u_2 v_2 \cos \theta_2 \right ]
+ w \left [ (h_3 u_3^2 + g {h_3^2 \over 2}) \sin \theta_3 + h_3 u_3 v_3 \cos \theta_3 \right ]  
= 0 .\ee

Using the branch velocities and neglecting the transverse components we get
$$ {g \over 2} \cos \theta_2 (|BC|~h_{BC}^2 -|DE| ~h_{DE}^2 )
+  {g \over 2} \cos \theta_3 (|EF|~h_{EF}^2 -|GH| ~h_{GH}^2 )$$
\be  \label{cpy2}
-w ~ \left [h_1 v_1^2 + g {{h_1}^2 \over 2} \right ]
+ w \sin \theta_2 \left [ h_2 {u_2^{\parallel}}^2 + g {h_2^2 \over 2} \right ]
+ w \sin \theta_3 \left [ h_3 {u_3^{\parallel}}^2 + g {h_3^2 \over 2} \right ] = 0 .\ee


\subsection{Momentum Flux for the T-Fork}

Consider now the T-geometry shown in the right panel of Figure~\ref{elbow2}. The calculations are simpler so that we used this geometry to validate the approach numerically. The general fork domain $\mathcal{F}$ can be reduced to the square $ADFIA$ by taking $\theta_2=\pi, \theta_3=0$ and $B \to C \to A, ~~G\to H\to I$. Then~the Equations (\ref{cpx})~and~(\ref{cpy}) reduce to
\begin{eqnarray}\label{tcpx}
-h_1 u_1 v_1 -( h_2 u_2^2 + g{h_2^2 \over 2}) + h_3 u_3^2 + g{h_3^2 \over 2}=0,  \\
\label{tcpy}
-( h_1 v_1^2  + g{h_1^2 \over 2})  - h_2 u_2 v_2 + g{h_{23}^2 \over 2} + h_3 u_3 v_3 =0 ,
\end{eqnarray}
where the term $h_{23}$ is 
\be\label{h23} h_{23}^2 \equiv {1\over w}\int_{DF} {h^2} ~ds .\ee

We will see that it can be obtained by interpolation of $h_2$ and $h_3$.


\subsection{Effective 1D Model for the T-Fork}

\begin{figure}
  \centering
  \vspace{1em}
  \resizebox{14 cm}{5 cm}{\includegraphics[angle=0]{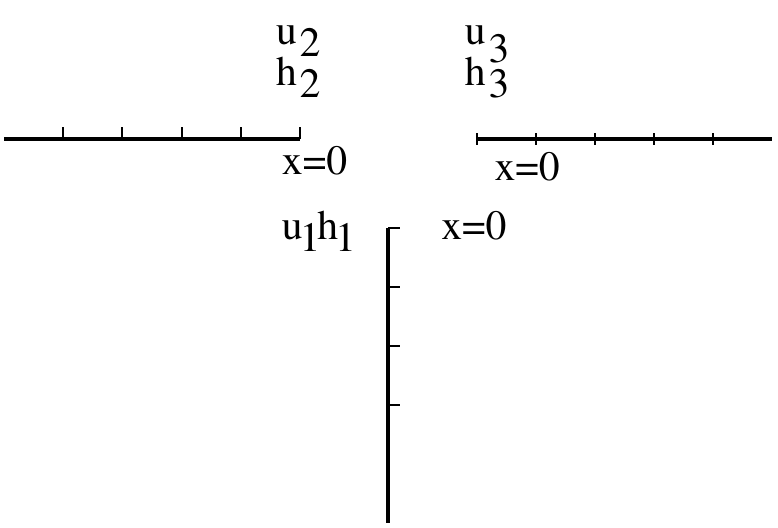}}
  \caption{Space discretization for the 1D approximation.}
  \label{discret}
\end{figure}

The pseudo-conservation laws (\ref{mass}), (\ref{energy}), (\ref{cpx2}) and (\ref{cpy2}) established in the previous section in the limit $w \to 0$ provide a formal connection between $h,{u^{\parallel}}$ in branches 1,2 and 3. In principle, they enable to approximate the 2D problem (\ref{sha1})--(\ref{sha2b}) by three 1D shallow water equations 
\begin{eqnarray}\label{sh1D}
  H^i_t + (H^i U^i)_x=0, \\
  (H^i U_i)_t + (H^i {U^i}^2 + {g H_i^2 \over 2})_x=0,
\end{eqnarray}
where $i=1,2,3$ correspond to the different branches. These 1D shallow water equations can be solved using a standard finite difference scheme, see for example~\cite{dkm11}. The~discretization is shown in Figure~\ref{discret} where the first nodes in each branch have values $H=h_i,U=u_i$.  The coupling equations between these three nodes given by (\ref{mass}),(\ref{cpx2}) and (\ref{energy}) would be solved using a Newton iteration.

\section{Numerical Solutions of the 2D Shallow Water Equations} \label{sec5}

The approximation described in the previous section holds if the error remains small. We now evaluate this error by solving numerically the 2D problem (\ref{sha1})--(\ref{sha2b}), compute $h,{u^{\parallel}}$ and see how these values agree with the pseudo-conservation laws (\ref{mass}),(\ref{cpx2}),(\ref{cpy2}) and~(\ref{energy}). We chose the T geometry shown in the right panel of Figure~\ref{elbow2} for simplicity and considered symmetric and non symmetric initial conditions. We also increased the wave amplitude to estimate the effect of the non linearity.

The Equations (\ref{sha1})--(\ref{sha2b}) were discretized using as space unit the depth $d$. The~time unit was $\sqrt{d \over g}$. The variables and fields was rescaled as 
\be\label{norm}
x' = {x \over d}, ~~t' = t \sqrt{g \over d}, ~~h'={x \over d}, ~~u'={u \over\sqrt{g d}}.\ee

This amounts to taking $d=1,~g=1$ in (\ref{sha1})--(\ref{sha2b}).

We solved the nonlinear shallow water equations using a first order finite volume scheme on an unstructured triangular mesh produced with the Gmsh meshing software (see details in~\cite{Volna}). We used the width $w=0.125$ and the typical size of the triangles is $0.02$. The time advance used a variable order Adams--Bashforth--Moulton multistep solver (implemented in Matlab under ode113 subroutine~\cite{Shampine1997}). The~relative and absolute tolerances were set to $10^{-5}$.

The initial condition is taken as a travelling solitary wave of velocity $c$. This is an exact solution for the mass conservation law. We used a solitary wave inspired by the Serre theory~\cite{Serre1}, (see~\cite{dcmm13} for the modern variational derivation)
\begin{eqnarray}\label{init}
h(x,y,t=0) = d + \eta(y),\\
v(x,y,t=0) =  c {  \eta(y)  \over d + \eta(y)}, \\
\eta(y)=  a~ {\rm sech}^2({1\over 2}k(y-y_0)),
\end{eqnarray}
where the speed is 
$$c = \sqrt{g(d+a)} .$$
The other parameters were 
$$ g=1,~~k=1,~~d=1,~~a=1.,~~x_0=y_0=2.5$$.

The wave was chosen so that its extension $ 2/k=2$ is much larger than the width $w=0.125$. Below we discuss the effect of the width.

The four pseudo-conservation laws for the mass, momenta and energy (\ref{mass},\ref{cpx2},\ref{cpx2},\ref{energy}) on the fork domain $ADFIA$ are
{\small
\begin{align} 
\label{dm} \delta m  \equiv  -h_1 v_1 -h_2 u_2 + h_3 u_3  = 0,~~\\
\label{dpx} \delta p_x   \equiv 
-h_1 u_1 v_1 -( h_2 u_2^2 + g{h_2^2 \over 2}) + h_3 u_3^2 + g{h_3^2 \over 2}= 0.~~\\
\label{dpy} \delta p_y   \equiv 
-( h_1 v_1^2  + g{h_1^2 \over 2})  - h_2 u_2 v_2 + g{h_{23}^2 \over 2} + h_3 u_3 v_3 = 0 ,~~\\
\label{de} \delta e   \equiv  -v_1 (g{h_1^2} + h_1 {v_1^2 \over 2}) 
-u_2 (g{h_2^2 } +h_2{u_2^2 \over 2} )
+ u_3 (g{h_3^2 } + h_3 {u_3^2 \over 2}) = 0,~~
\end{align}}
where we introduced the residuals $\delta m,~\delta p_x,~\delta p_y$ and $\delta e$.

Two situations were considered. We considered a symmetric situation where the wave is incident from branch 1 and a non symmetric situation where the wave was send into the fork from branch 3. In both cases, the number of unknowns was the same; see Table~\ref{tab2}.

\begin{table}
\centering
\caption{The two different dynamic problems for the T-branch.}
\label{tab2}
\begin{tabular}{ccc}
   \toprule
   \textbf{Type}     &  \textbf{Known}    &  \textbf{Unknown}   \\ \midrule
    wave in branch 1 &   $h_1,v_1$   &   $h_2,u_2,h_3,u_3$ \\ 
    wave in branch 3 &   $h_3,u_3$   &   $h_1,v_1,h_2,u_2$ \\ 
   \bottomrule
\end{tabular}
\end{table}

The wave mass and wave energy in each branch have been calculated. They are defined as
$$ M_w = \int_\Omega (h-d)  ~dx dy,$$
$$ E_w = \int_\Omega  {1 \over 2} \left [ g (h-d)^2 + (u^2  + v^2 )h \right ]  ~dx dy . $$

Energy will propagate very differently in problems 1 and 2. In the next sections we examine in detail the two types of problems and use the conservation laws  to establish jump conditions for the 1D effective model.

To verify the approximation given by the relations (\ref{dm})--(\ref{de}), we also computed the time evolution of the quantities $h_1,h_2,h_3,v_1,u_2,u_3$ from the 2D direct numerical simulations.  We used a scattered linear interpolation to estimate these physical variables along the four different segments of the fork region from the unstructured triangular mesh data.


\subsection{Wave Incident into Branch 1}

\subsubsection{Small Amplitude Waves $a/d=0.1$}

Consider the wave mass, at $t=0$: $M_1^0 =57~ 10^{-3}, ~~~~  M_2^0 =M_3^0 =0$. After the wave has passed, at $t=6.5,~~~~ M_1=0,~~~~ M_2=M_3= 26 $. We have $2\times 26 = 52$ which shows the conservation of mass. Notice the depression in branch 1 after wave passes. Almost all energy is transferred to branches 2 and~3.

\begin{figure}
  \centering
  \resizebox{12 cm}{5 cm}{\includegraphics{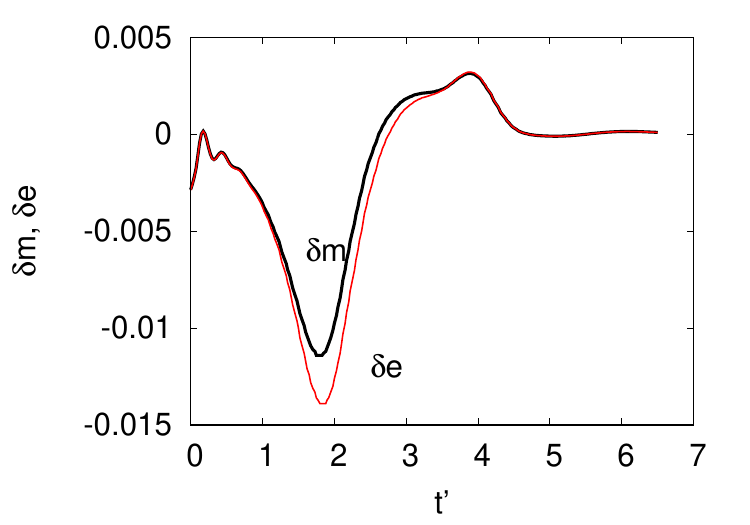}}
  \caption{Time evolution of the mass and energy quantities $\delta m, ~\delta e $ for $a/d=0.1$.}
  \label{c1dmde_01}
\end{figure}

Here our balance laws hold well for both the mass and the energy,
see {Figure}~\ref{c1dmde_01}.

\begin{figure}
  \centering
  \resizebox{15 cm}{7 cm}{\includegraphics[angle=0]{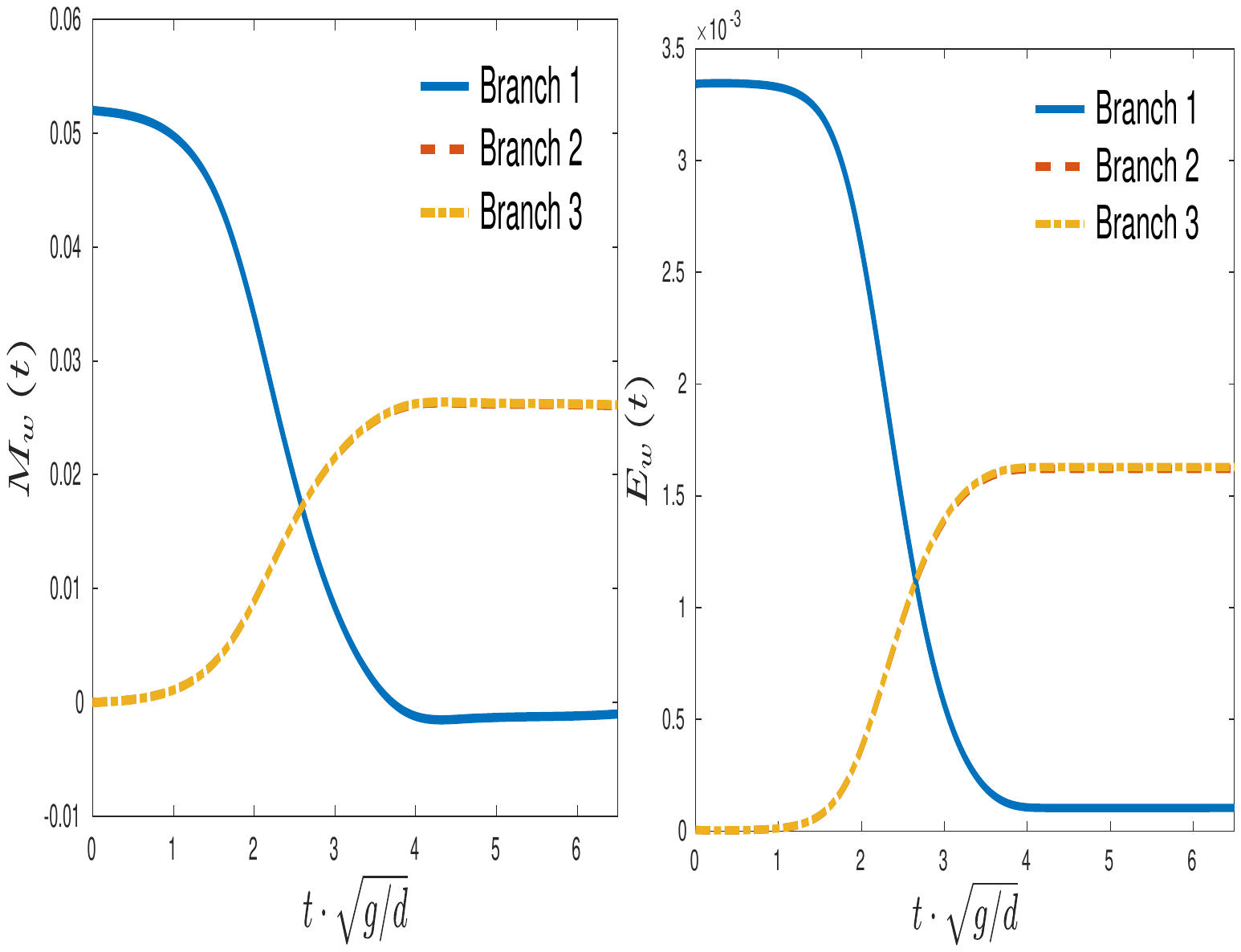}}
  \caption{Time evolution of the wave mass $M_w$ (\textbf{left}) and the wave energy $E_w$ (\textbf{right}) for a wave incident in branch 1 for $a/d=0.1$.}
\label{c1me_01}
\end{figure}

\begin{figure}
  \centering
  \resizebox{15 cm}{8 cm}{\includegraphics[angle=0]{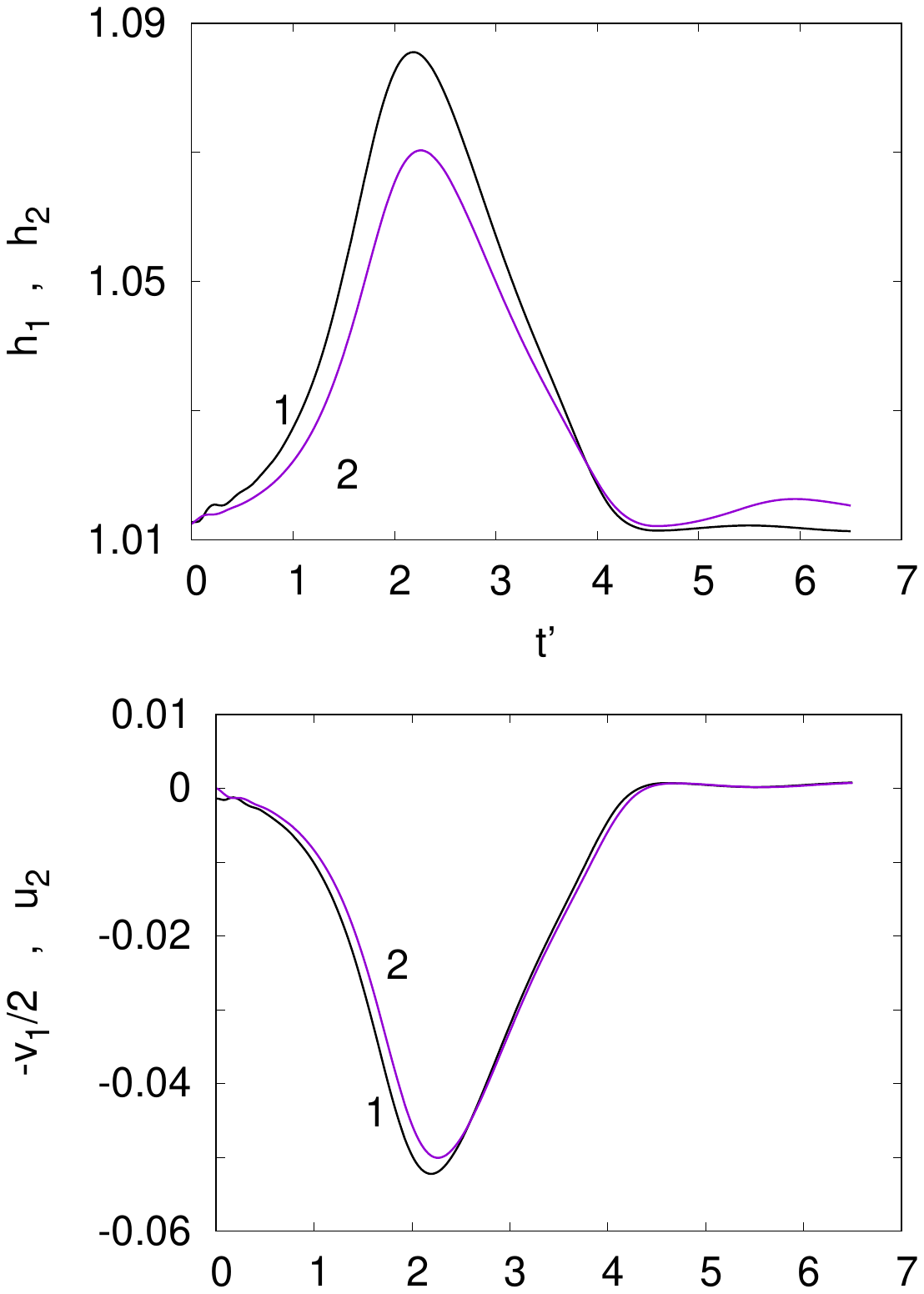}}
  \caption{Time evolution of $h_1$, $h_2$ (\textbf{top}) and $v_1/2,u_2$ (\textbf{bottom}) for $a/d=0.1$.}
\label{c1h12u12_01}
\end{figure}

We can use them to obtain $u_2, h_2$. Assume symmetry $h_2=h_3 , u_2=-u_3$. The balance laws reduce~to
\begin{eqnarray}
  -h_1 v_1 -2 h_2 u_2 =0 , \\
  -v_1 (g h_1^2  + h_1 v_1^2/2) -2 u_2 (g h_2^2  + h_2 u_2^2/2)=0.
\end{eqnarray}
Since $v_1^2 , u_2^2 \ll gh^2$ we can neglect the terms $v_1^2, u_2^2$ of the second equation. The resulting relations are satisfied by 
\be \label{stoker} h_2=h_1, ~  u_2=-v_1/2 , \ee
which are the Stoker conditions. These are in good agreement with the simulations as shown by Figure~\ref{c1h12u12_01}.

\subsubsection{ Very Large Amplitude Waves $a/d=2$}

\begin{figure}
  \centering
  \resizebox{12 cm}{5 cm}{\includegraphics{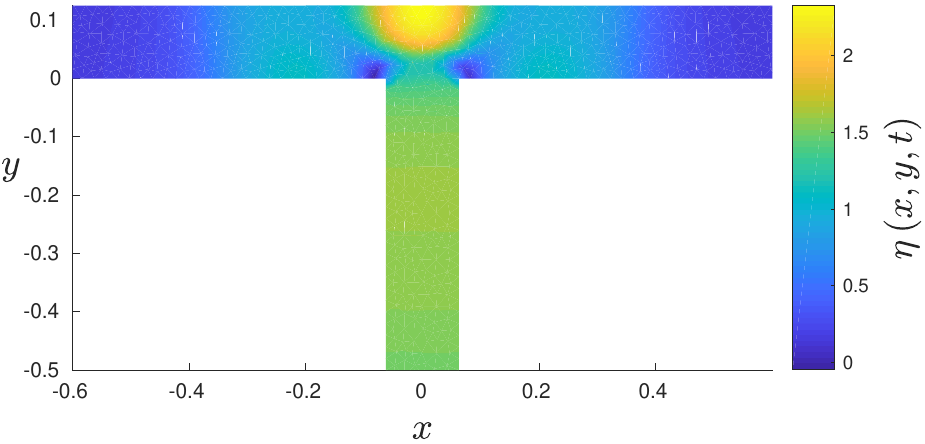}}
  \caption{Snapshot of the surface elevation $h$ at time $t=0.9$ for a wave incident in branch 1 for $a/d=2$.}
\label{photo1}
\end{figure}

In this case, 2D effects start to appear. Figure~\ref{photo1} shows a snapshot of the surface elevation $h$ for a wave such that $a/d=2$. Notice the lump $h \approx 2$ on the edge of the domain.

Despite the evidence of 2D effects, the overall transfer of wave mass and
wave energy from branch 1 to branches 2 and 3 does not vary significantly
as $a/d$ changes from $0.1$ to $2$.

\begin{figure}
  \centering
  \resizebox{15 cm}{6 cm}{\includegraphics[angle=0]{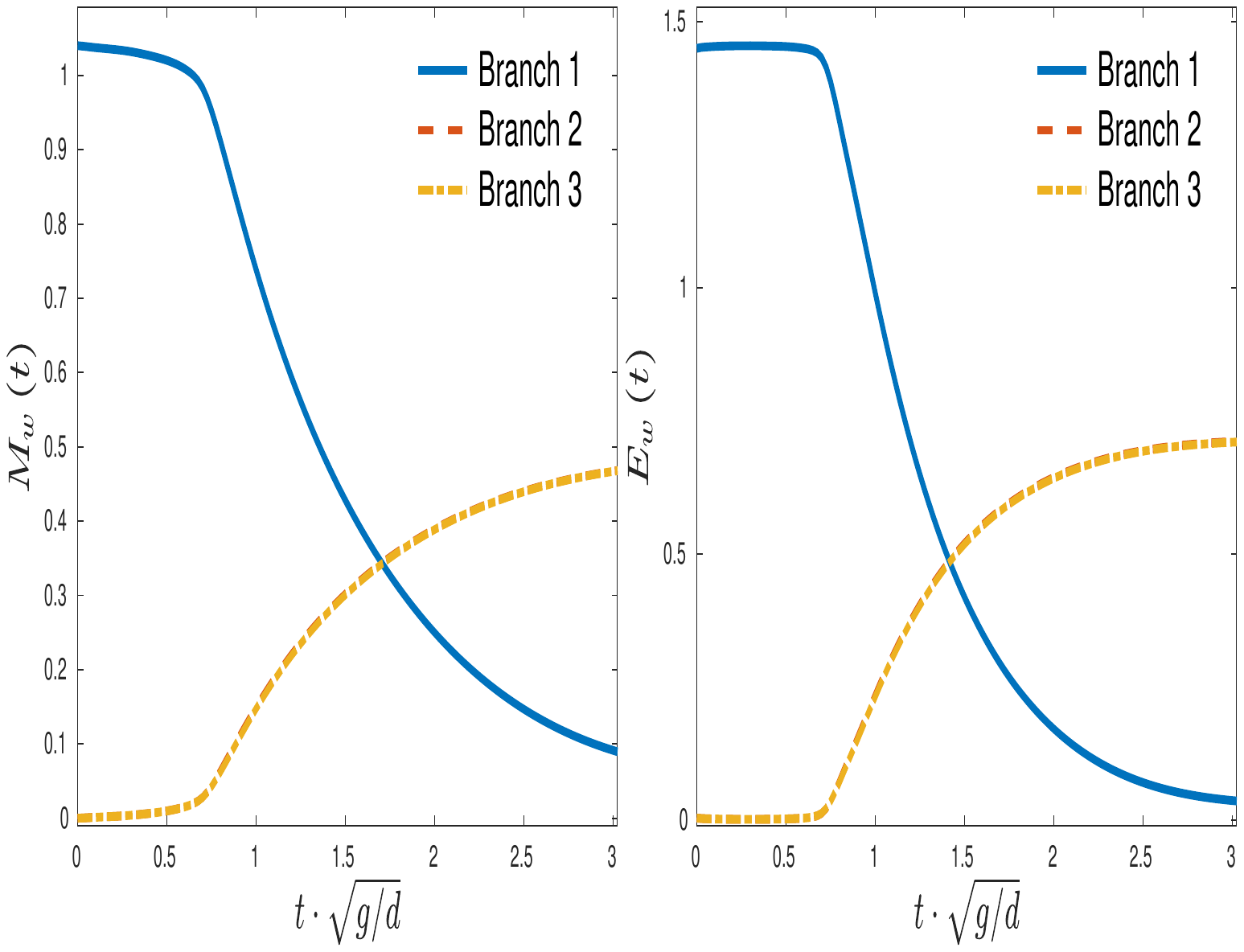}}
  \caption{Time evolution of the wave mass $M_w$ (\textbf{left}) and the wave energy $E_w$ (\textbf{right}) for a wave incident in branch 1 for $a/d=2$.}
  \label{c1me_2}
\end{figure}

\begin{figure}
  \centering
  \resizebox{12 cm}{5 cm}{\includegraphics{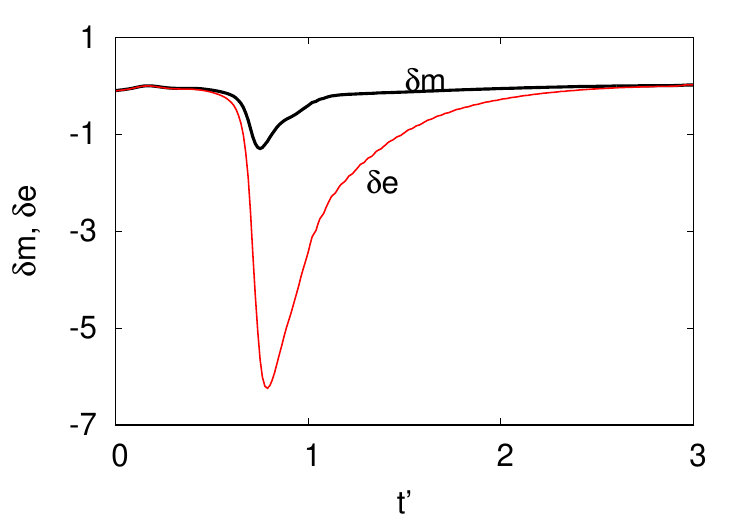}}
  \caption{Time evolution of the mass and energy quantities $\delta m, ~\delta e $ for $a/d=2$.}
  \label{c1dmde_2}
\end{figure}
Notice that the mass relation is better satisfied than the energy relation.

\begin{figure}
  \centering
  \resizebox{14 cm}{6.5 cm}{\includegraphics[angle=0]{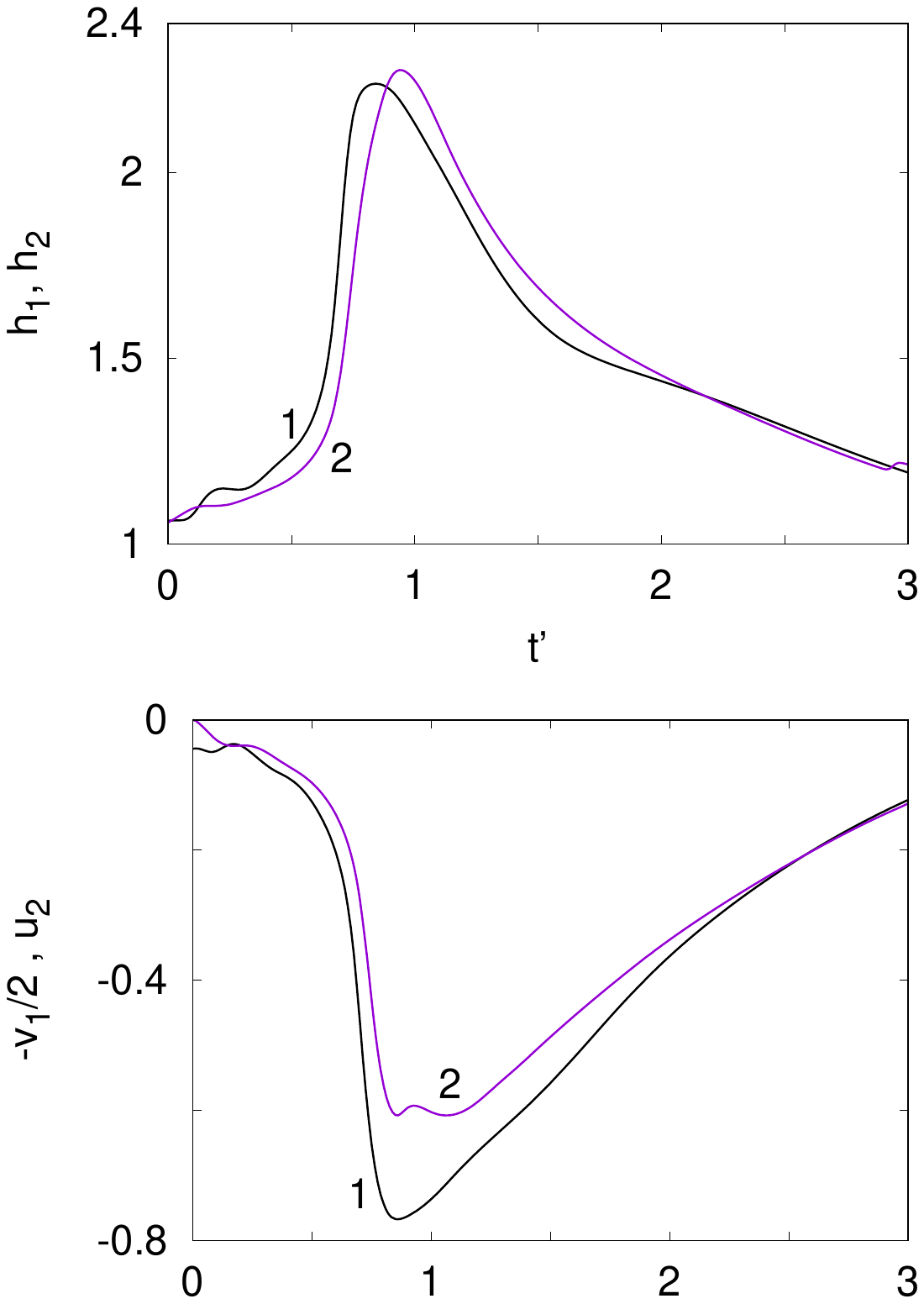}}
  \caption{Time evolution of $h_1$, $h_2$ (\textbf{top}) and $v_1/2$, $u_2$ (\textbf{bottom}) for $a/d=2$.}
  \label{c1h12u12_2}
\end{figure}

Again the Stoker relations (\ref{stoker}) give a good approximation as shown by Figure~\ref{c1h12u12_2} which show that $h_2 \approx h_1 $ and $u_2 \approx v_1/2$. The~price to pay to approximate the 2D situation by a 1D effective model is an energy loss at the junction.

Also remark that for the approximation to hold it is crucial that the wave be wider than $w$ and not too fast. If these conditions are not met, $h_2$ and $u_2$ will be delayed from $h_1, v_1$ and will need to describe what happens in the fork. We observed this for a larger channel $w=1$ and the same parameters.


\subsection{Wave Incident into Branch 3}

\begin{figure}
  \centering
  \resizebox{15 cm}{9 cm}{\includegraphics{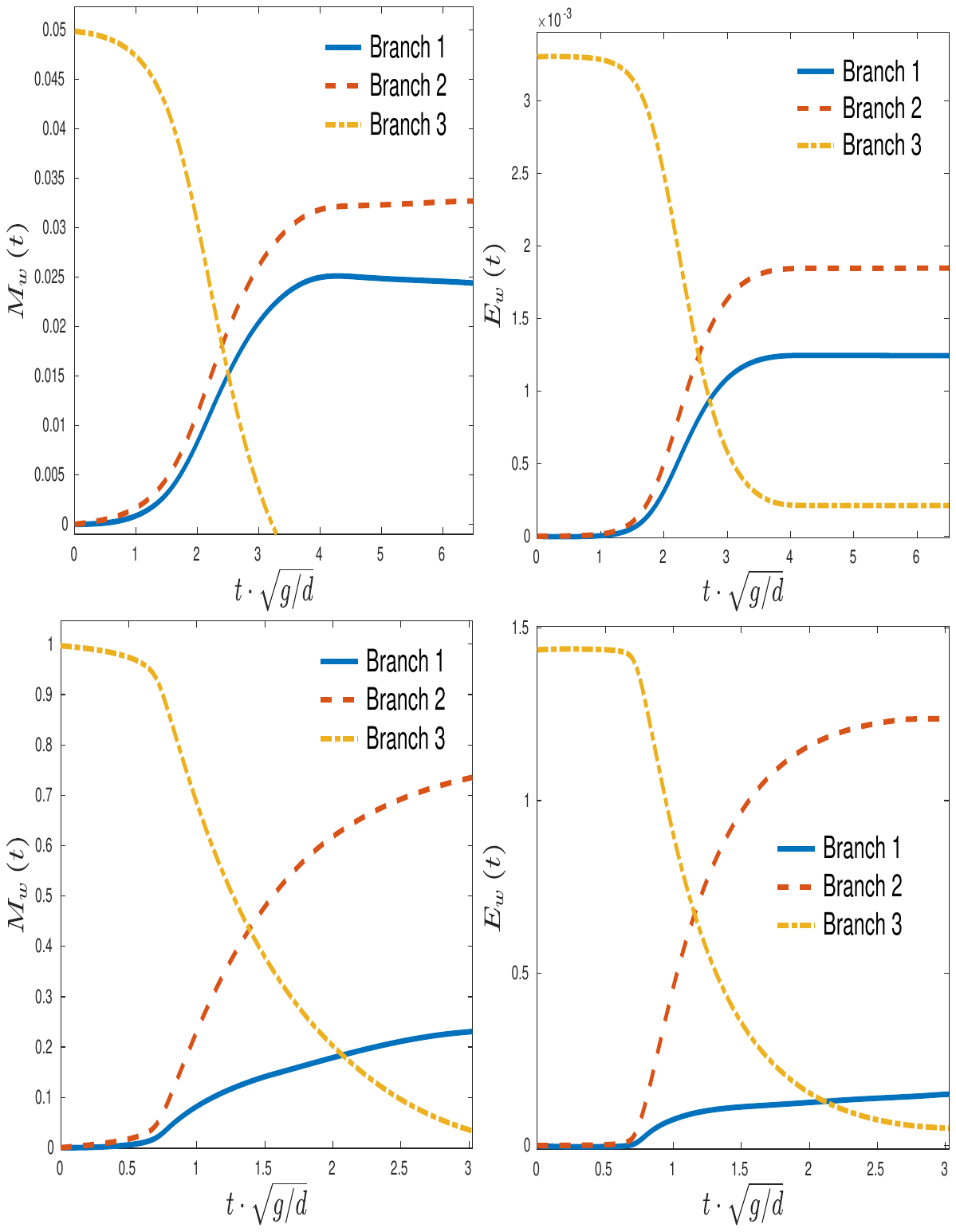}}
  \caption{Time evolution of the wave mass $M_w$ (\textbf{left}) and the~wave energy $E_w$ (\textbf{right}) for a wave incident in branch 3 for $a/d=0.1$ (\textbf{top panels}) and $a/d=2$ (\textbf{bottom panels}). Notice the~different scales.}
  \label{me2_012}
\end{figure}

For this configuration, we observe a significant difference in behavior as the wave amplitude increases. Figure~\ref{me2_012} shows the time evolution of the wave mass and wave energy for $a/d=0.1$ (top panels) and $a/d=2$  (bottom panels). Small amplitude waves get transmitted to branch 1 as much as to branch 2. On the other hand, large amplitude waves are predominantly transmitted to branch 2. The~mass entering branch 2 is three times larger than the one entering branch 1; for energies, the factor is six.

\subsubsection{Small Amplitude Waves $a/d=0.1$}

\begin{figure}
  \centering
  \resizebox{15 cm}{6 cm}{\includegraphics{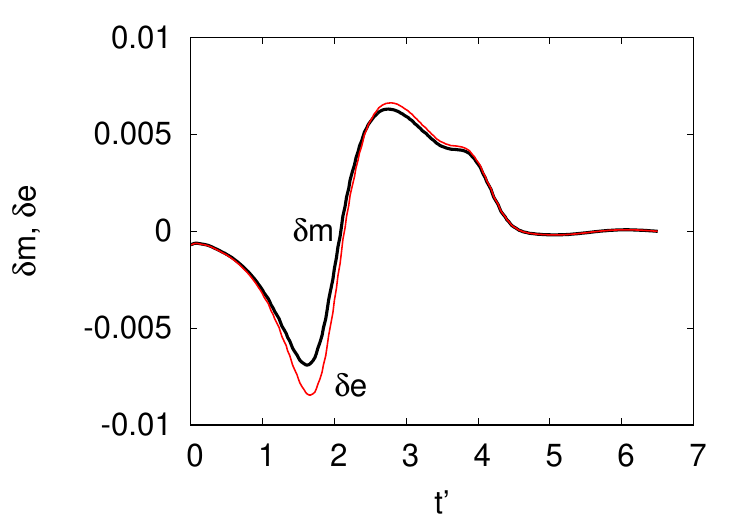}}
  \caption{Time evolution of the mass and energy quantities $\delta m, ~\delta e $ for $a/d=0.1$.}
  \label{c2dmde_01}
\end{figure}

\begin{figure}
  \centering
  \resizebox{16 cm}{7 cm}{\includegraphics{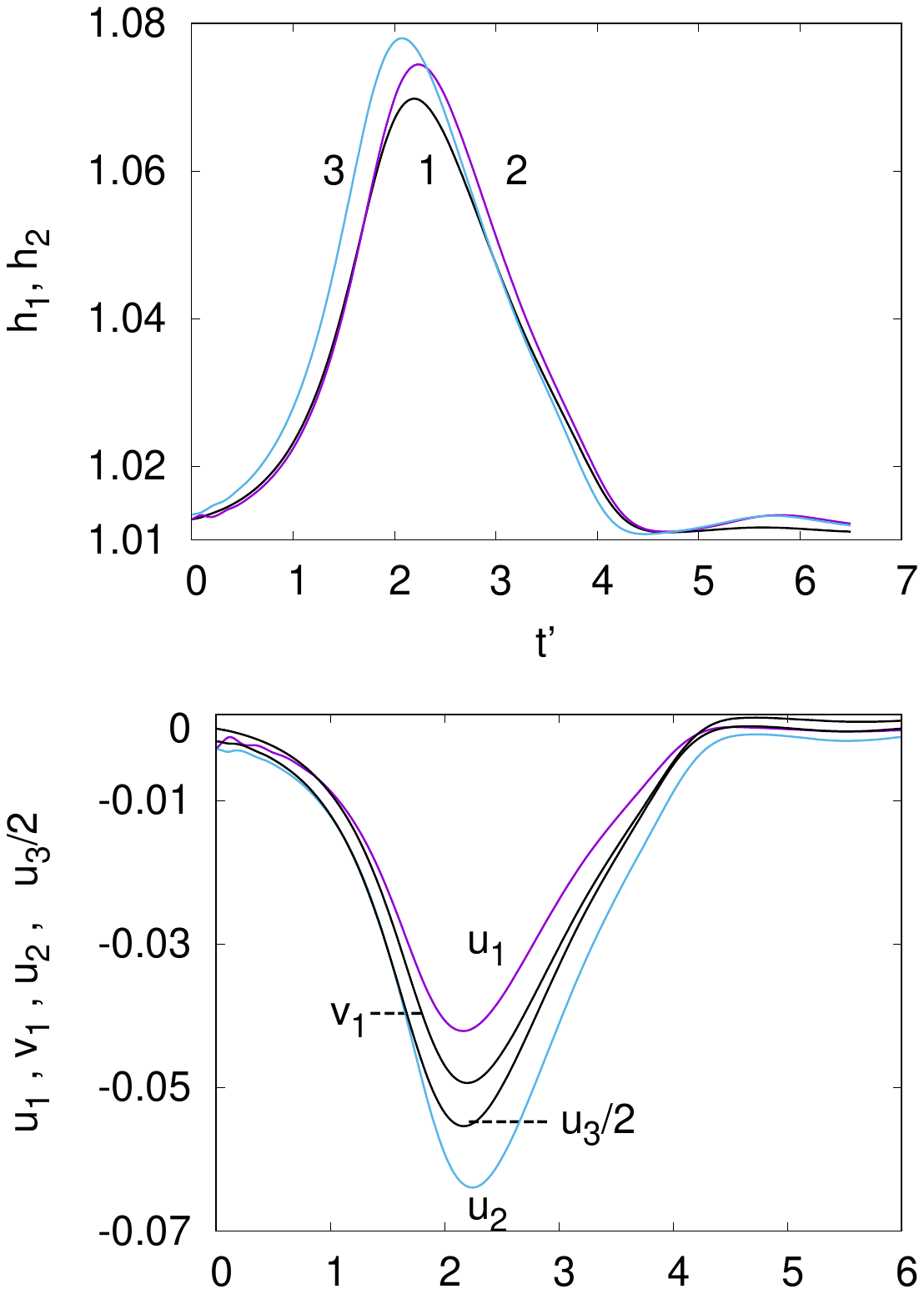}}
  \caption{Time evolution of $h_1,h_2$ (top) and $u_2,u_1,v_1$ (bottom) for $a/d=0.1$.}
  \label{c2h12u12_01}
\end{figure}

First observe that $u_1$ is non zero and close to $v_1$. Nevertheless, the mass and energy residuals $\delta m$ and $\delta e$ are small as seen in Figure~\ref{c2dmde_01}. The wave elevation $h$ does not vary much from one branch to the other as seen in the top panel of Figure~\ref{c2h12u12_01}. The~velocities $u_2$ and $v_1$ verify $u_2 \approx u_3/2, ~~v_1\approx u_3/2$. These two results show that the Stoker conditions hold for this small amplitude.

\subsubsection{Large Amplitude Waves $a/d=1$}

\begin{figure}
  \centering
  \resizebox{12 cm}{5 cm}{\includegraphics{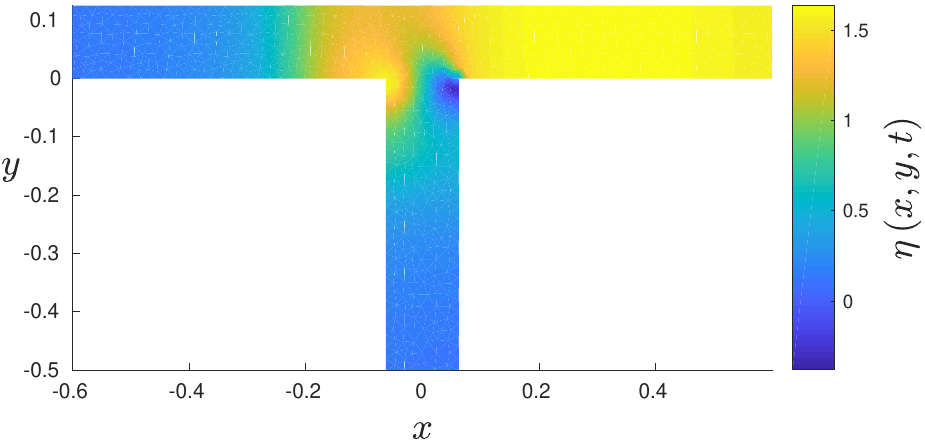}}
  \caption{Snapshot of the surface elevation $h$ at time $t=0.8$ for a wave incident in branch 3 for $a/d=2$.}
  \label{photo2}
\end{figure}

Figure~\ref{photo2} shows $h(t=0.8)$ for a wave incident in branch 3 for $a/d=2$. Notice the complex structure of the flow at the junction. There is some recirculation so that the flow is essentially 2D and not amenable to a 1D reduction.

\begin{figure}
  \centering
  \resizebox{11 cm}{4.5 cm}{\includegraphics{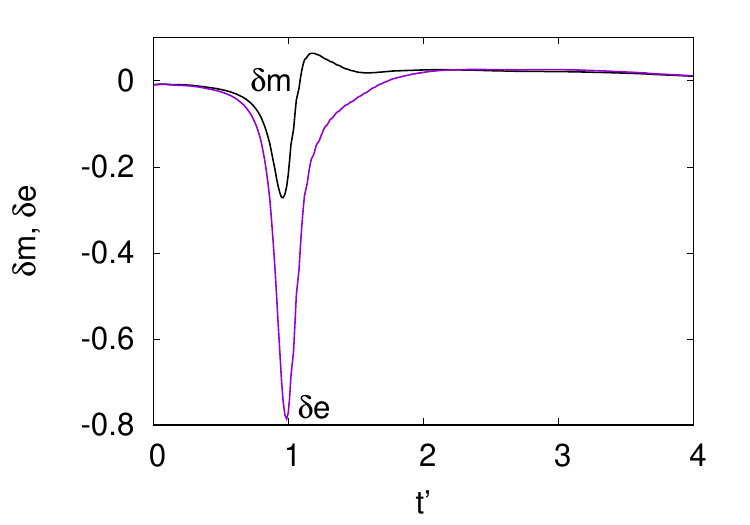}}
  \caption{Time evolution of the mass and energy quantities $\delta m, ~\delta e $ for $a/d=2$.}
  \label{c2dmde_1}
\end{figure}

Nevertheless, for a smaller amplitude $a/d=1$, the balance laws (\ref{dm})--(\ref{de}) give some insight into the flow. Figure~\ref{c2dmde_1} shows the mass $\delta m$ and energy $\delta e$. The mass is much better conserved than the~energy.

\begin{figure}
  \centering
  \resizebox{12 cm}{4.4 cm}{\includegraphics{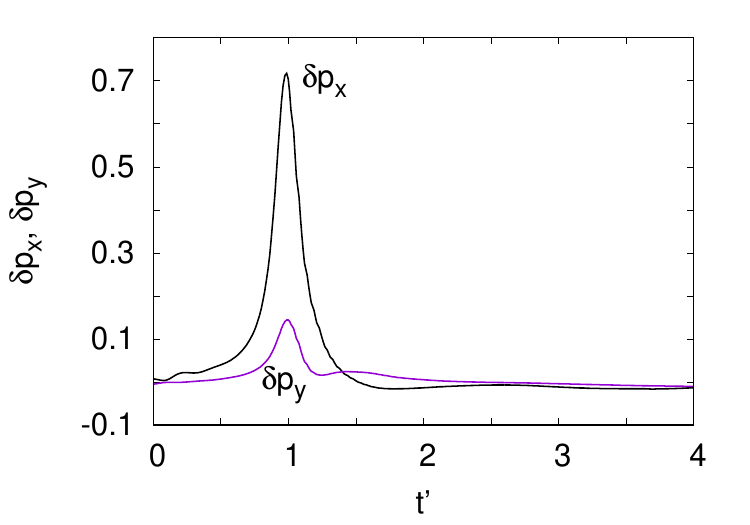}}
  \caption{Time evolution of the $x$ and $y$ momenta quantities $\delta p_x, ~\delta p_y $ for $a/d=1$.}
  \label{c2dpxy_1}
\end{figure}

The momenta (\ref{dpx}), (\ref{dpy}) are plotted in Figure~\ref{c2dpxy_1}.

\begin{figure}
  \centering
  \resizebox{15 cm}{7 cm}{ \includegraphics{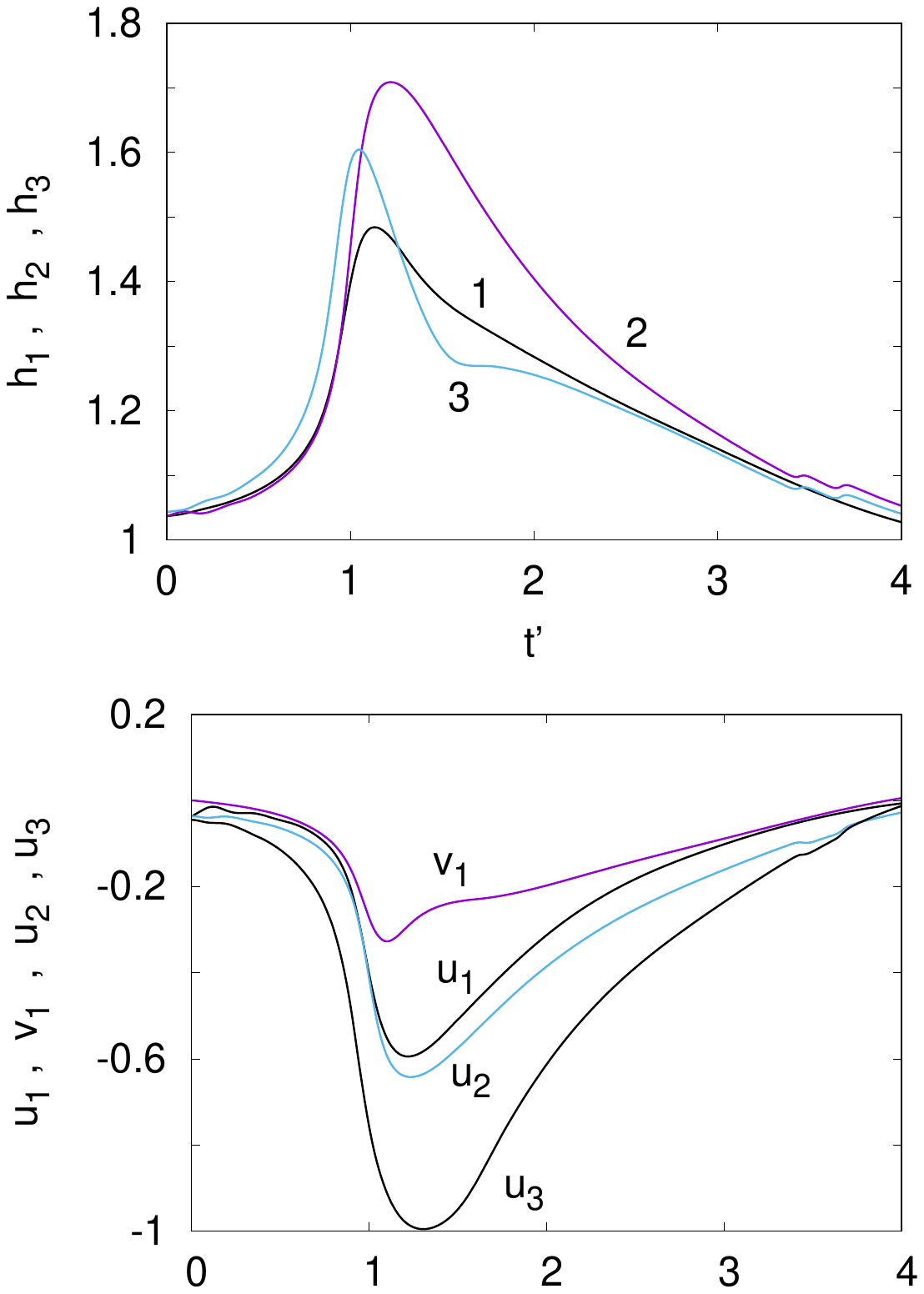}}
  \caption{Time evolution of $h_1,h_2$ (\textbf{top}) and $u_2,u_1,v_1$ (\textbf{bottom}) for $a/d=1$.}
  \label{c2h12u12_1}
\end{figure}

When the wave is coming from branch 3, an obvious solution is
\be\label{sol_pb2} v_1=0,~~u_2=u_3,~~h_2=h_3,~~h_1=h_2. \ee
This is simplistic, in reality $v_1 \neq 0$ but remains small. The~horizontal component $u_1$ is non zero and close to $u_2$ as shown in Figure~\ref{c2h12u12_1}.

\begin{figure}
  \centering
  \resizebox{15 cm}{6 cm}{\includegraphics[angle=0]{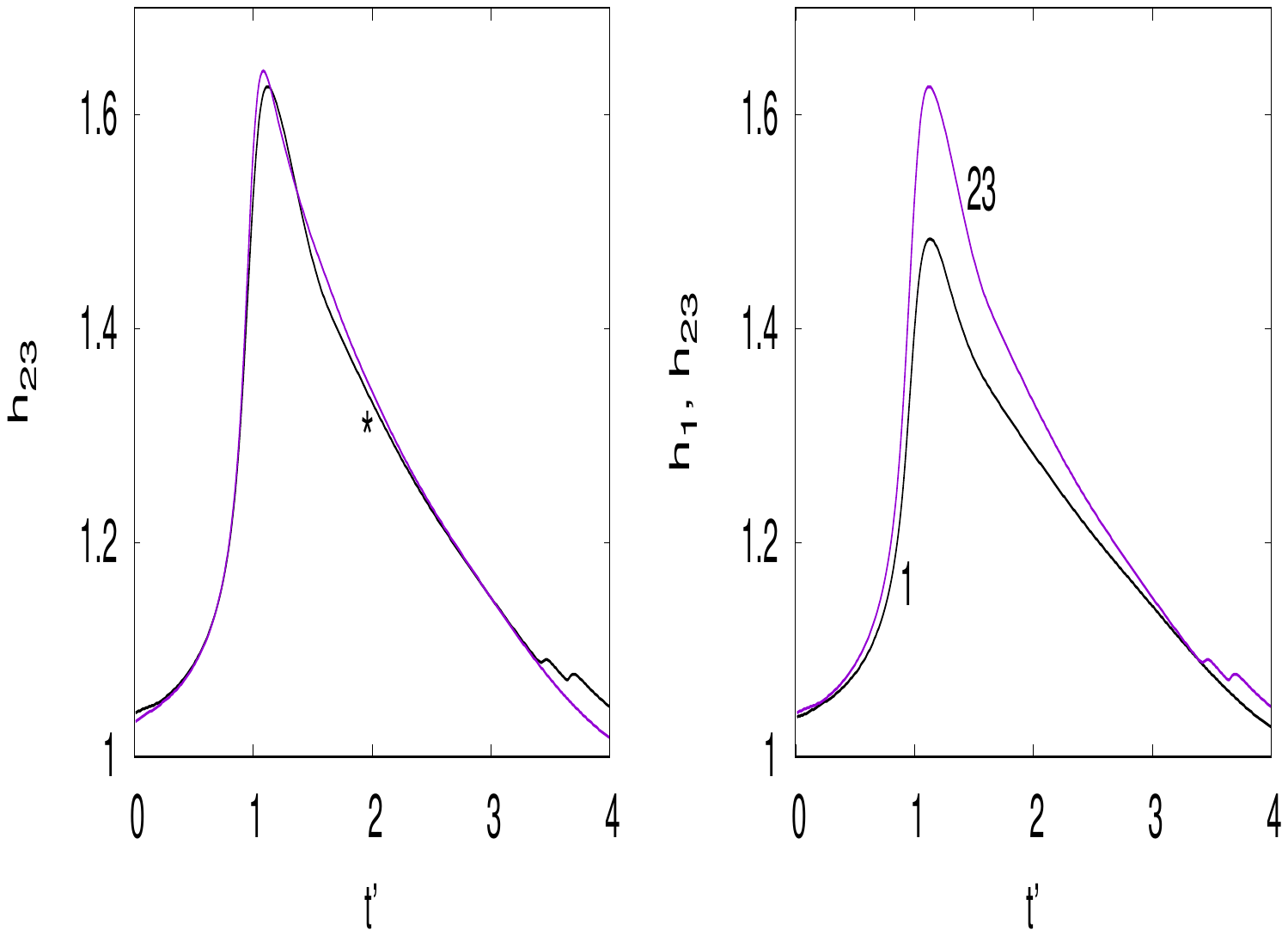}}
  \caption{(\textbf{Left}) panel, time evolution of the quantity $h_{23}$ from (\ref{h23}) together with the approximation (\ref{h23i}) indicated by the $*$ symbol. (\textbf{Right}) panel, time evolution of $h_{23}$ and $h_1$.}
  \label{c2h123_1}
\end{figure}

The mass equation and $y$ momentum equations allow to extract relations between $v_1, h_1,h_2, u_2,v_1, h_3,u_3$. Assuming $v_1,u_2,u_3$ smaller than $h_1^2, h_2^2, h_3^2$, we have 
\begin{eqnarray} \label{sys_2}
  v_1 = {h_3 u_3  - h_2 u_2  \over h_1}, \\
  h_1 =h_{23}. 
\end{eqnarray}
The quantity (\ref{h23}) in the $y$ component of the momentum is computed from the numerical solution. It is plotted as a function of time together with the estimate
\be \label{h23i} h_{23}^i = \sqrt{ {1\over 2 } (h_2^2 + h_3^2)},\ee 
in the left panel of Figure~\ref{c2h123_1}. As can be seen, the agreement is very good.

\begin{figure}
  \centering
  \resizebox{15 cm}{6 cm}{\includegraphics[angle=0]{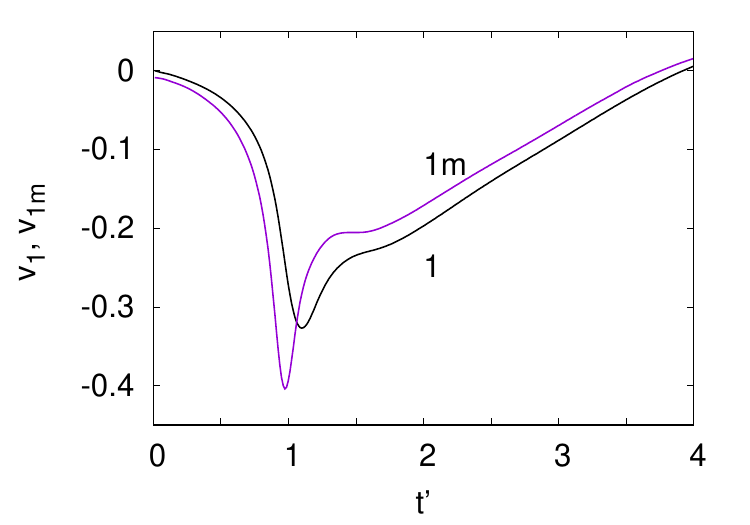}}
  \caption{Time evolution of the quantity $v_{1m}$ the mass conservation law \eqref{sys_2} and $v_1$ from the 2D numerical solution.}
  \label{c2v11m_1}
\end{figure}

The velocity $v_{1m}$ given by the mass conservation relation agrees semi-quantitatively with the value $v_1$ estimated from the 2D numerical solution. Both quantities are plotted as a function of time in Figure~\ref{c2v11m_1}. Note the delay due to the time the wave needs to propagate from one interface to the other. The $y$ momentum conservation law is not satisfied so that there is no additional equation to estimate~$u_2$.

\section{Discussion and Conclusions}\label{sec6}

The results of the previous section show that for large amplitudes and an asymmetric fork Stoker's interface conditions do not hold and the angle of the fork plays a role. This seems to contradict the findings of Shi et al.~\cite{shi05}. Two reasons show that there is no contradiction. First, the amplitude of our waves ($a/d \approx 1$) are much larger than the ones presented in~\cite{shi05} ($a/d \approx 0.3$) so that nonlinear effects are much stronger in our study. The~other point is that the ${\rm sech}^2$ initial condition is an exact solution of the Boussinesq equations, but not of the nonlinear shallow water equations. For the Boussinesq equations, we also expect an angle dependence, even for narrow channels, when the amplitude becomes large. To~see this, we examine the reduction of the equations for a fork.

The Boussinesq equations read
\begin{eqnarray}
\label{bou1} h_t + \nabla \cdot [(1+h)\nabla \varphi] =0, \\
  \label{bou2} \varphi_t + {1 \over 2} (\nabla \varphi)^2 +h 
  -{1 \over 3} (\Delta \varphi)_t=0 ,
\end{eqnarray}
where $h(x,y,t)$ is the water elevation. The velocity potential $\varphi(x,y,t)$ is such that $(u,v)^T=\nabla \varphi$. The~boundary conditions are non slip $\nabla \varphi \cdot \mathbf{n}=0$. Integrating the equations on the fork domain $\mathcal{F}$ (left panel of Figure~\ref{elbow2}) we get
\begin{eqnarray}
\label{ibou1}
\partial_t \int_{\mathcal{F}} h dx dy
- \int_{IA \cup CD \cup FG}  (1+h)\nabla \varphi \cdot \mathbf{n} ds,\\
\label{ibou2}
\partial_t \int_{\mathcal{F}} (\varphi-{1 \over 3}\Delta \varphi)dx dy
  + \int_{\mathcal{F}} ({1 \over 2} (\nabla \varphi)^2 +h) dx dy=0 .
\end{eqnarray}
Neglecting the time evolution in the fork region, we get the following 
interface conditions
\begin{eqnarray}\label{int_bou1} 
  (1+h_1)u_1^{\parallel} + (1+h_2)u_2^{\parallel}+ (1+h_3)u_3^{\parallel}=0, \\
  \label{int_bou2}
  \int_{\mathcal{F}} {1 \over 2}\left [ (\nabla \varphi)^2 +h \right ] dx dy=0.
\end{eqnarray}
Note how the first equation reduces to Kirchhoff's law for small $h$. The~second equation contains is an integral over the whole domain and depends on the angle of the fork. For small angles, we can assume that $\nabla \varphi = u^{\parallel}$ so that the conditions reduce to
\begin{eqnarray}\label{rint_bou1}
  (1+h_1)u_1^{\parallel} + (1+h_2)u_2^{\parallel}+ (1+h_3)u_3^{\parallel}=0, \\
  \label{rint_bou2} {1 \over 2}(u_1^{\parallel})^2 +h_1 
  + {1 \over 2}(u_2^{\parallel})^2 +h_2 
  + {1 \over 2}(u_3^{\parallel})^2 +h_3 =0.
\end{eqnarray}
Not surprisingly, these conditions are very close to the ones obtained by Nachbin and Simoes~\cite{ns15}, except for the Jacobian of the conformal transformation.

To conclude, we studied the propagation of shallow water waves in a fork between three narrow channels. We considered both the 2D numerical solution and a homothetic reduction procedure that gives coupling conditions at the interface. For such narrow widths, the delay experienced by the wave is negligible so that one can envision describing the junction by an effective 1D PDE model. 

Our reduction enabled us to derive balance laws for the mass, momenta and energy of the flow across a general junction. For small amplitude waves, these laws reduce to the commonly used Stoker jump conditions, giving these a formal justification. We verified these Stoker conditions on the 2D numerical solutions of the shallow water equations for symmetric and non symmetric conditions. Then, the angle of the junction does not play any role. This happens also for a general nonlinear wave equation; we had seen this a previous study for the particular case of the sine-Gordon equation~\cite{cd14}.

For large amplitude shallow water waves, the situation depends on the symmetry of the fork. For~a symmetric fork, the Stoker conditions are approximately verified. This is explained by the strong constraint imposed by the symmetry. Then,~the only solution of the balance laws corresponds to the Stoker conditions. When the fork is non symmetric as in our case 2, more information is needed about what happens inside the fork. The~quantities ${u_i^{\parallel}}, ~i=1,2,3$ are velocities projected in the direction of the branches and this projection leads to a loss of information. Far from the junction, the flow is quasi-1D so that not much is lost. On the contrary, inside the junction, the flow is full 2D. A possible solution, to be studied in the future would be to use the full conservation law including the time dependent term. Then we would introduce a fictitious node inside the junction and couple it to the boundaries using average differential equations obtained by integrating (\ref{sha1})--(\ref{sha2b}) on the fork domain.


\bigskip
\subsection*{Acknowledgments}
\addcontentsline{toc}{subsection}{Acknowledgments}

The authors thank Tim \textsc{Minzoni} and Nick \textsc{Ercolani} for useful discussions. We also thank the Centre R\'egional Informatique et d'Applications Num\'eriques de Normandie for the use of its computers. J.-G. ~\textsc{Caputo} was partially supported by ANR grant ``Fractal grid''. D.~\textsc{Dutykh} acknowledges the support of CNRS through the project PEPS InPhyNiTi ``FARA''.
 
\bigskip



\bigskip\bigskip

\end{document}